\documentclass[english,3p]{elsarticle}
\usepackage[T1]{fontenc}
\usepackage[latin9]{inputenc}
\usepackage{amsmath}
\usepackage{amssymb}
\usepackage{graphicx}

\makeatletter
\usepackage{babel}

\makeatother

\usepackage{babel}
\begin{document}

\title{The dynamics of the pendulum suspended on the forced Duffing oscillator}

\author[rvt]{P. Brzeski}

\author[rvt]{P. Perlikowski\corref{cor1}}

\ead{przemyslaw.perlikowski@p.lodz.pl}

\author[rvt1]{S. Yanchuk}

\author[rvt]{T. Kapitaniak}

\cortext[cor1]{Corresponding author}

\address[rvt]{Division of Dynamics, Technical University of Lodz, Stefanowskiego
1/15, 90-924 Lodz, Poland}

\address[rvt1]{Institute of Mathematics, Humboldt University of Berlin, Unter den
Linden 6, 10099 Berlin, Germany}

\address{{*}przemyslaw.perlikowski@p.lodz.pl}
\begin{abstract}
We investigate the dynamics of the pendulum suspended on the forced
Duffing oscillator. The detailed bifurcation analysis in two parameter
space (amplitude and frequency of excitation) which presents both
oscillating and rotating periodic solutions of the pendulum has been
performed. We identify the areas with low number of coexisting attractors
in the parameter space as the coexistence of different attractors
has a significant impact on the practical usage of the proposed system
as a tuned mass absorber. \end{abstract}
\maketitle

\section{Introduction}

The dynamics of the pendulum or systems containing the pendulum is
probably one of the oldest scientific topics. Simple parametrically
excited pendulum shows extremely complex behaviour \cite{Leven198171,1985PhyD}.
Miles \cite{Miles:1988:RSB:51797.51803} shows that for the pendulum
the route to chaos leads through symmetry breaking pitchfork bifurcation
and cascades of period doubling bifurcations. Comprehensive analytical
investigation of the pendulum with different forcing was presented
by Bryant and Miles \cite{Bryant1990,Bryant1990_part1,Bryant1990_part2}.
In the forced system one can consider two main parameters: the amplitude
and the frequency of excitation. Such a bifurcation diagram for parametrically
forced pendulum with horizontally moving point of the suspension was
presented by Bishop and Clifford for periodic oscillations (PO) \cite{Bishop19961537}
and for periodic rotations (PR) \cite{Clifford1995191}. In \cite{Horton2011436}
bifurcation analysis was extended to elliptic movement of suspension
point both for PO and PR. The analytical investigation of oscillating
and rotating motions of pendulum was done using averaging, small parameter,
harmonic balance, and other methods\cite{Xu2007311,Xu20051537,Banerjee199621,Bishop19961537,Kholostova1995553,Kim19976613,kimLee1995,Kobes2000,Trueba2003911}.
This analysis allows understanding of the pendulum dynamics in the
neighbourhood of the locked periodic solutions. The interesting effect
can be observed when the symmetry of the pendulum is broken, i.e.,
the imperfection term is added to the potential function \cite{Sofroniou2006673}.
In such an asymmetric system the symmetry breaking pitchfork bifurcation
disappears and the sudden decrease of the amplitude of the PO at the
first period doubling bifurcation can be regarded as a precursor of
an escape or problems with system's behaviour.

The crucial point in modelling of such systems is a good approximation
of damping coefficients (viscous and frictional damping). There are
well known methods for linear \cite{Liang2005,Liang2006} systems
but as far as the pendulum is oscillating with large amplitude, the
linear approximation does not give sufficient results. This problem
has been solved by Xu et al. \cite{Horton2007,Xu2007172}. They show
an efficient method to extract the damping coefficients (viscous damping
and dry friction) from time series of freely oscillating pendulum
and consider the influence of the shaker (the source of the forcing).

Most previous works on the dynamics of the pendulum suspended on the
forced oscillator consider the linear oscillators. Such a system can
be considered as a modification of the classical tuned mass absorber
\cite{Tondl2006,springerlink:10.3103/S105261880804002X}. Early works
\cite{hatwal:657,hatwal:663} give approximate results by the method
of harmonic balance in the primary parametric instability zone, which
allows calculation of the separate regions of stable and unstable
harmonic solutions. Further analysis allows understanding of the dynamics
around primary and secondary resonances \cite{Bajaj1994,Balthazar20011075,Cartmell1994173,kecik2005,Song2003747}.
In the recent work Ikeda proposed the usage of two pendulums mounted
in the same pivot as a tuned mass absorber \cite{ikeda:011012}. His
experimental results show good agreement with the numerical simulations
and his model can be considered as a good alternative to one pendulum
on the pivot in the design of tuned mass absorbers. 

A good understanding of dynamics of tuned mass absorber with linear
base system gives possibility to extend investigation to systems with
non--linear base. Non--linearity in considered class of systems is usually
introduced by changing the linear spring into non--linear one \cite{Warminski2009612,WARMINSKI2001363}
or magnetorheological damper \cite{ISI:000289102700001}. In a few
papers on this topic one can find an analytical study of the dynamics
of Duffing -- pendulum systems around principal and secondary resonances
\cite{kecik2005,Warminski2009612,WARMINSKI2001363,BVG2008,4723450}.
The main conclusion coming from the above mentioned papers is that
non-linear spring in the base system causes enlargement of parameters
range where pendulum can be used as a tune mass absorber.

In this paper we consider the pendulum suspended on the forced non--linear
Duffing oscillator. The purpose of our analysis is to study the emergence
and the stability of PO and PR in two parameters space: the amplitude
and the frequency of excitation. We identify the regions with one
stable periodic solution, several coexisting periodic solutions, quasi-periodicity,
and chaotic behaviour.

The paper is organized as follows. In Section 2 we formulate the dimensionless
equations of motion. The possible scenarios of pendulum\textquoteright{}s
destabilization are presented in Section 3. Section 4 shows two-dimensional
bifurcation diagrams for PO, PR, as well as one-parameter continuations
for representative values of parameters. The influence of non--linearity
of spring on absorbing properties of the pendulum is studied as well.
In Section 5 we show the regions in two-dimensional parameter space
where one, two, or several coexisted attractors can be observed.
Finally, in Section 6 we summarize our results.

\section{Model of the system}

\begin{figure}
\begin{centering}
\includegraphics{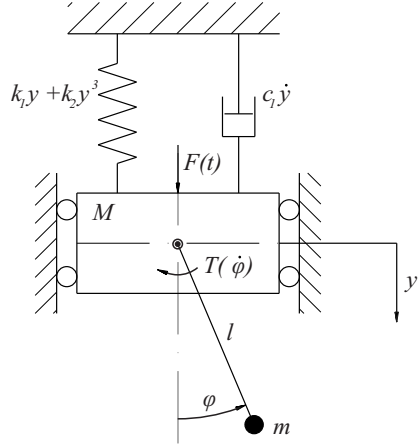} 
\par\end{centering}

\caption{Model of system.\label{fig:Model-of-system}}
\end{figure}

The analyzed system is shown in Fig. \ref{fig:Model-of-system}. It
consists of a Duffing oscillator with a suspended pendulum. The Duffing
system is forced by periodical excitation and moving in a vertical
direction. The position of mass $M$ is given by coordinate $y$ and
the angular displacement of pendulum (position of the mass $m$) is
given by angle $\varphi$. The equations of motion can be derived
using Lagrange equations of the second type. The kinetic energy $T$,
potential energy $V$, and Rayleigh dissipation $D$ are given respectively
by the following equations: 
\begin{equation}
T=\frac{1}{2}(M+m)\dot{y}^{2}-ml\dot{y}\dot{\varphi}\sin\varphi+\frac{1}{2}ml^{2}\dot{\varphi}^{2},\label{eq:ener_kin}
\end{equation}

\begin{equation}
V=\frac{1}{2}k_{1}y^{2}+\frac{1}{4}k_{2}y^{4}+mgl(1-\cos\varphi),\label{eq:ener_pot}
\end{equation}

\begin{equation}
D=\frac{1}{2}c_{1}\dot{y}^{2},\label{eq:dyssypacja}
\end{equation}
where $M$ is mass of the Duffing oscillator, $m$ is mass of the
pendulum, $l$ is length of the pendulum, $k_{1}$ and $k_{2}$ are
linear and non--linear parts of spring stiffness, and $c_{1}$ is a
viscous damping coefficient of the Duffing oscillator.  The generalized
forces are given by the following formula:

\begin{equation}
Q=F\left(t\right)\frac{{\partial y}}{\partial y}+Tq\left(\dot{{\varphi}}\right)\frac{{\partial\varphi}}{\partial\varphi},\label{eq:si=00003D000142y_uog}
\end{equation}
 where $F\left(t\right)=F_{0}\cos\nu t$ is a periodically varying
excitation with amplitude $F$ and frequency $\nu$, $Tq\left(\dot{\varphi}\right)=c_{2}\dot{\varphi}$
is a damping torque with damping coefficient $c_{2}$. The damper
of pendulum is located in a pivot of the pendulum (not shown in Fig.
\ref{fig:Model-of-system}). The damping in the pivot of pendulum is composed of viscous and dry friction damping \cite{Horton2007}. Here we neglect dry friction component (to have a continuous system) and assume small value of viscous part ($1 \%$ of critical damping).

One can derive two coupled second order differential equations:

\begin{equation}
(M+m)\ddot{y}-ml\ddot{\varphi}\sin\varphi-ml\dot{\varphi}^{2}\cos\varphi+k_{1}y+k_{2}y^{3}+c_{1}\dot{y}=F_{0}\cos\nu t,\label{eq:wym_masa}
\end{equation}

\begin{equation}
ml^{2}\ddot{\varphi}-ml\ddot{y}\sin\varphi+mlg\sin\varphi+c_{2}\dot{\varphi}=0.\label{eq:wym_wahadlo}
\end{equation}
 In the numerical calculations we use the following values of Duffing
oscillator's parameters: $M=5.0\,\mathrm{[kg]}$, $k_{1}=162.0\,\mathrm{\left[\frac{N}{m}\right]}$,
$k_{2}=502.0\,\mathrm{\left[\frac{N}{m}\right]}$, $c_{1}=3.9\,\mathrm{\left[\frac{Ns}{m}\right]}$
and the following values of the pendulum's parameters: $m=0.5\,\mathrm{[kg]}$,
$l=0.1\,\mathrm{[m]}$, $c_{2}=0.001\,\mathrm{\left[Nms\right]}$.
We neglect static deflection of mass $M$.

Introducing dimensionless time $\tau=t\omega_{1}$, where $\omega_{1}^{2}=\frac{k_{1}}{M+m}$
is natural linear frequency of Duffing oscillator, we reach dimensionless
equations:

\begin{equation}
\begin{array}{c}
\ddot{x}-ab\ddot{\gamma}\sin\gamma-ab\dot{\gamma}^{2}\cos\gamma+x+\alpha x^{3}+d_{1}\dot{x}=f\cos\mu\tau,\\
\\
\ddot{\gamma}-\frac{1}{b}\ddot{x}\sin\gamma+\sin\gamma+d_{2}\dot{\gamma}=0,
\end{array}\label{eq:row bez}
\end{equation}
 where $\omega_{2}^{2}=\frac{g}{l}$, $a=\frac{m}{M+m}$, $b=\left(\frac{\omega_{2}}{\omega_{1}}\right)^{2}$,
$\alpha=\frac{k_{2}l^{2}}{(M+m)\omega_{1}^{2}}$, $f=\frac{F_{0}}{(M+m)l\omega_{1}^{2}},$
$d_{1}=\frac{c_{1}}{(M+m)\omega_{1}}$, $d_{2}=\frac{c_{2}}{ml^{2}\omega_{2}}$,
$\mu=\frac{\nu}{\omega_{1}}$, $x=\frac{y}{l}$, $\dot{x}=\frac{\dot{y}}{\omega_{1}l}$,
$\ddot{x}=\frac{\ddot{y}}{\omega_{1}^{2}l}$, $\gamma=\varphi,$ $\dot{\gamma}=\frac{\dot{\varphi}}{\omega_{2}},$
$\ddot{\gamma}=\frac{\ddot{\varphi}}{\omega_{2}^{2}}$.

The dimensionless parameters of the system have the following values:
$a=0.091$, $b=3.33$, $\alpha=0.031$, $d_{1}=0.132$ and $d_{2}=0.02$.
Both subsystems (Duffing oscillator and the pendulum) have linear
resonance for $\mu=1.0$, so around this value we expect the appearance
of the complex dynamics. Amplitude $f$ and frequency $\mu$ of the
excitation are taken as control parameters.

\section{Destabilization of pendulum}

System (\ref{eq:row bez}) possesses three qualitatively different
regimes. The first one is the regime when the oscillations of the
Duffing system are not large enough to destabilize the pendulum. Hence,
the pendulum is at stable steady state $\gamma=0$. The second regime
appears when pendulum destabilizes and starts oscillating. The third
regime is characterized by the rotating motions of the pendulum. The
transition from the stable quiescence state $\gamma=0$ to oscillations
can be understood from the theoretical point of view as the destabilization
of the invariant manifold $\gamma=0$. Indeed, the manifold $\gamma=\dot{\gamma}=0$
is invariant with respect to (\ref{eq:row bez}) and the dynamics
on the manifold is described by the single Duffing oscillator: 
\begin{equation}
\ddot{x}_{0}+x_{0}+\alpha x_{0}^{3}+d_{1}\dot{x}_{0}=f\cos\mu\tau,
\end{equation}
 with an effective mass $m+M$. The linear stability of motions on
this manifold is given by the variational equation:
\begin{equation}
\delta\ddot{\gamma}+d_{2}\delta\dot{\gamma}+\left(1-\frac{1}{b}\ddot{x}_{0}(\tau)\right)\delta\gamma=0,
\end{equation}
which has form of a linear system with respect to the variation $\delta\gamma$,
perturbed periodically by the Duffing $x_{0}(\tau)$. Such a parametrically
perturbed system is known to possess destabilization regions (parametric
resonance) when the frequency of the perturbation $x_{0}(\tau)$ is
rationally related to the frequency of the perturbation. Since in
resonances zones the Duffing oscillator is usually locked $1:1$ with
the external frequency, we expect the oscillation regions in the parameter
space close to $\mu\approx2$ (the most prominent resonance) as well
as $\mu\approx1/2,\,1,\,2/3$, etc. 

\begin{figure}
\begin{centering}
\includegraphics{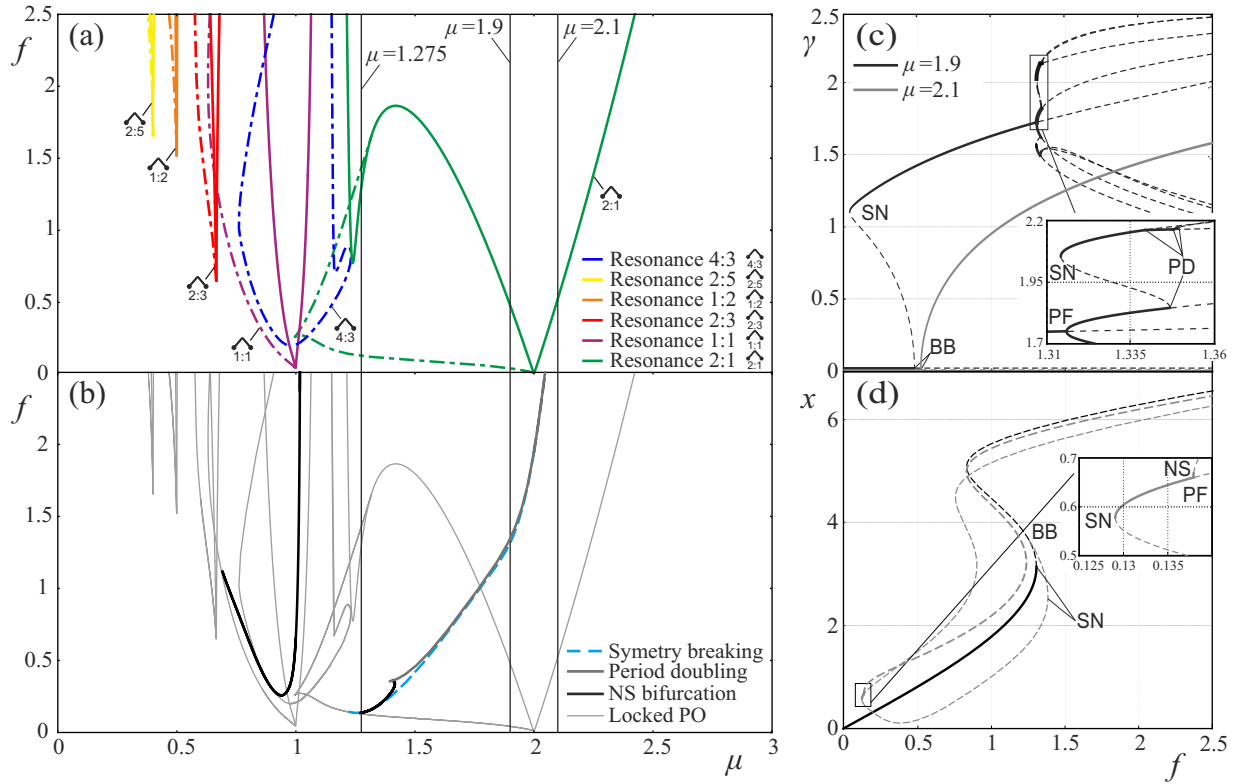} 
\par\end{centering}

\caption{(colour online) Bifurcation diagrams of the PO of the pendulum for system
(\ref{eq:row bez}) in two parameter space ($f$, $\mu$). (a) Tongues
of resonances: solid lines correspond to the destabilization of the
pendulum and birth of PO, dashed-dotted lines denote saddle-node bifurcations
of PO. Different colours distinguish PO with different locking ratios
between the pendulum and the excitation. (b) Bifurcation lines where
the destabilization of PO occurs via symmetry breaking, period doubling,
or Neimark-Saker bifurcations. (c) One-dimensional bifurcation diagrams
for $\mu=1.9$ (black lines) and $\mu=2.1$ (grey lines) shows a route
to $2:1$ resonance in more details. (d) One-dimensional bifurcation
diagram ($\mu=1.275$) for $2:1$ tongue, black line shows PO with
pendulum in hanging down position and grey one with oscillating pendulum.
In (c,d) solid and dashed lines correspond to stable and unstable
PO respectively. Abbreviations: BB - branching bifurcation, NS - Neimark-Saker
bifurcation, PF - symmetry breaking pitchfork bifurcation, SN - saddle-node
bifurcation. \label{fig:bif_2para} }
\end{figure}

As a result of the destabilization of pendulum, PO appear, where $x_{0}(\tau)$
is locked $1:1$ to the external force and $\gamma(\tau)$ to some
other ratio depending on the resonance tongue. Since after the destabilization
of the pendulum, the emerged periodic solution still coexists with
the unstable solution $(\gamma=0,x_{0}(\tau))$, we will call it branching
bifurcation. In Section \ref{sec:Two-parameters-continuation} we
develop a two-dimensional bifurcation diagram with respect to $f$ and $\mu$. Branching bifurcations are shown as solid lines on this
bifurcation diagram in Figs. \ref{fig:bif_2para}(a,b) delineating
the resonance tongues.

\section{Two parameters continuation\label{sec:Two-parameters-continuation}}

In this section we present two bifurcation diagrams calculated in
two-parameter space: amplitude $f$ versus frequency $\mu$ of excitation.
We focus our attention on bifurcations of the pendulum. Duffing system,
due to excitation, is oscillating in the whole considered range of
parameters. Such plots give an overview of system dynamics showing
the most important periodic solutions, i.e., periodic solutions with
significant area of existence. Our calculations have been performed
using software for numerical continuation Auto07p \cite{autp_doedel}.
As the starting points in our calculations, we use the steady state
with $f=0.0$ and follow it for different values of $\mu$ detecting
the bifurcations leading to different periodic motions. Moreover,
in a few cases we start from periodic orbits calculated by the direct
integration of eq. (\ref{eq:row bez}). For integration we use the
fourth order Runge-Kutta method. The stability of periodic solutions
is given by the set of Floquet multipliers \cite{Kuznetsov1995}.

\subsection{Oscillatory solutions in plane $(f,\mu)$}

In Fig. \ref{fig:bif_2para}(a) we show main resonances for which
the pendulum is oscillating. Different colours of bifurcation lines
indicate the borders of different resonances tongues, i.e., areas
with different locking ratio between the pendulum and the excitation
frequency. Duffing oscillator after branching bifurcation is always
locked $1:1$ with excitation frequency, further bifurcations can
change this ratio. The natural dimensionless frequency of the pendulum
and the Duffing oscillator are equal to one. We find the resonances
with the following locking rations: $1:1$ (purple line), $1:2$ (orange
line), $2:1$ (green line), $4:3$ (blue line), $2:5$ (yellow line),
and $2:3$ (red line). These borders of the resonance tongues are
of two kinds. Continuous lines correspond to the destabilization of
the pendulum at the branching bifurcation and the appearance of PO
with oscillating pendulum, and dashed-dotted lines to saddle-node
bifurcations of PO.

Resonance $4:3$ appears in the saddle-node bifurcation and it is
stable in a large range of parameters. Other resonances have qualitatively
similar structure with three bifurcation curves meeting in one point:
two solid lines and one dashed-dotted. The main resonant tongues ($1:1$
and $2:1$) come very close to the axis $f=0$ due to small friction.
We illustrate the corresponding bifurcation scenarios in Fig. \ref{fig:bif_2para}(c)
using one-dimensional bifurcation diagrams for the case of $2:1$
resonance. Fig. \ref{fig:bif_2para}(c) shows maximum amplitude
of the pendulum versus $f$ for $f\in(0.0,\,2.5)$ with fixed $\mu=1.9$
and $\mu=2.1$. Solid and dashed lines correspond to stable and unstable
PO respectively. The black line indicates PO calculated for $\mu=1.9$.
As it is easy to see, till $f=0.479$ the pendulum is in hanging down
position and only the Duffing system is oscillating. Then through
the subcritical branching bifurcation PO (which corresponds to the
oscillations of the pendulum) appears, but this branch of the PO is
unstable (continuous line on the left side of the edge of the tongue
(Fig. \ref{fig:bif_2para}(a) indicate this bifurcation). For $f=0.043$
the saddle-node bifurcation of PO takes place and this branch of PO
stabilizes. For $f=1.316$ one can observe the symmetry breaking pitchfork
bifurcation generating two asymmetric solutions. Both branches of
PO originated from pitchfork bifurcation, destabilize in the subcritical
period doubling bifurcations, in which the new unstable PO with doubled
period is born. This branch of PO stabilizes in the saddle-node bifurcation.
Especially interesting is the situation shown in the enlargement in
Fig. \ref{fig:bif_2para}(c). One can see the coexistence of the
$4:1$ asymmetric PO (both Duffing and pendulum are quadruple) first
with symmetric $2:1$ PO and then after symmetry breaking pitchfork bifurcation
of $2:1$ resonance with two asymmetric $2:1$ PO (after this bifurcation
Duffing oscillator is locked $2:1$ with excitation - the same ratio
as pendulum). Further increase of $f$ for $4:1$ asymmetric PO leads
through the period doubling scenario to chaos. This bifurcation route
shows that scenario described by Miles \cite{Miles:1988:RSB:51797.51803,Miles:1989:RSB:64936.64957}
in the considered case of system (\ref{eq:row bez}) becomes more
complicated.

In the same plot we present scenario for $\mu=2.1$ (grey line) where
the lower equilibrium position of the pendulum is destabilized by
the supercritical branching bifurcation for $f=0.524$ (bifurcation
takes place at continuous line on the right side of the edge of the
tongue (see Fig. \ref{fig:bif_2para}(a))) and the stability of
this solution does not change in the considered range of the amplitude
of excitation $f$.

The same scenarios, with division into right (only continuous line)
and left (continuous and dashed-dotted lines) sides of the edge of
the tongues are observed for other resonances (see Fig. \ref{fig:bif_2para}(a)).
In case of $2:1$ resonance branching bifurcation is a period doubling
bifurcation while for other resonances branching bifurcation leads
to different locking ratios. 

For $2:1$ resonance in the range $\mu\in(1.23,\,1.317)$ the continuous
green line is below the dashed-dotted green line. In this area we
observe the bifurcation scenario which is shown in Fig. \ref{fig:bif_2para}(d)
for $\mu=1.275$ in the range $f\in(0.0,\,2.5)$ versus maximum amplitude
of Duffing oscillator. The black solid line shows the growth of the
amplitude of the Duffing oscillations in the case when the pendulum
is in the lower equilibrium position. This PO loses its stability
in the saddle-node bifurcation (the pendulum persists in the equilibrium
position). Then we observe the branching bifurcation of the unstable
PO and the appearance of new branch of the PO for which the pendulum
is in $2:1$ resonance with excitation (grey line). After two saddle-node
bifurcations the branch of the PO stabilizes for $\mu=0.129$ (see
zoom in Fig. \ref{fig:bif_2para}d). Finally, for $\mu=0.1378$ the
PO loses its symmetry in the pitchfork bifurcation and through the
Neimark-Saker bifurcation ($\mu=0.1379$) becomes unstable. 

In Fig. \ref{fig:bif_2para}(b) we show main destabilization scenarios
for locked PO. The resonances tongues are marked by grey lines in
the background of the plot. The Neimark-Saker bifurcations (long curve
in Fig. \ref{fig:bif_2para}(b)) destroys the $1:1$ resonant PO,
i.e., above line of this bifurcation locked $1:1$ PO does not exist.
Other lines are connected to stability of $2:1$ tongue. The dashed
light blue line indicates the symmetry breaking pitchfork bifurcation
and just after it we observe the period doubling bifurcation (detailed
route to chaos is shown in zoom in Fig. \ref{fig:bif_2para}(c)).
At the end of period doubling line we detect the Neimark-Saker bifurcation,
which merges with the symmetry breaking pitchfork bifurcation and
$2:1$ resonance curve. Other PO presented in Fig. \ref{fig:bif_2para}(a)
are stable in the whole covering range but they are accessible only
for carefully chosen initial conditions.

\begin{figure}
\begin{centering}
\includegraphics{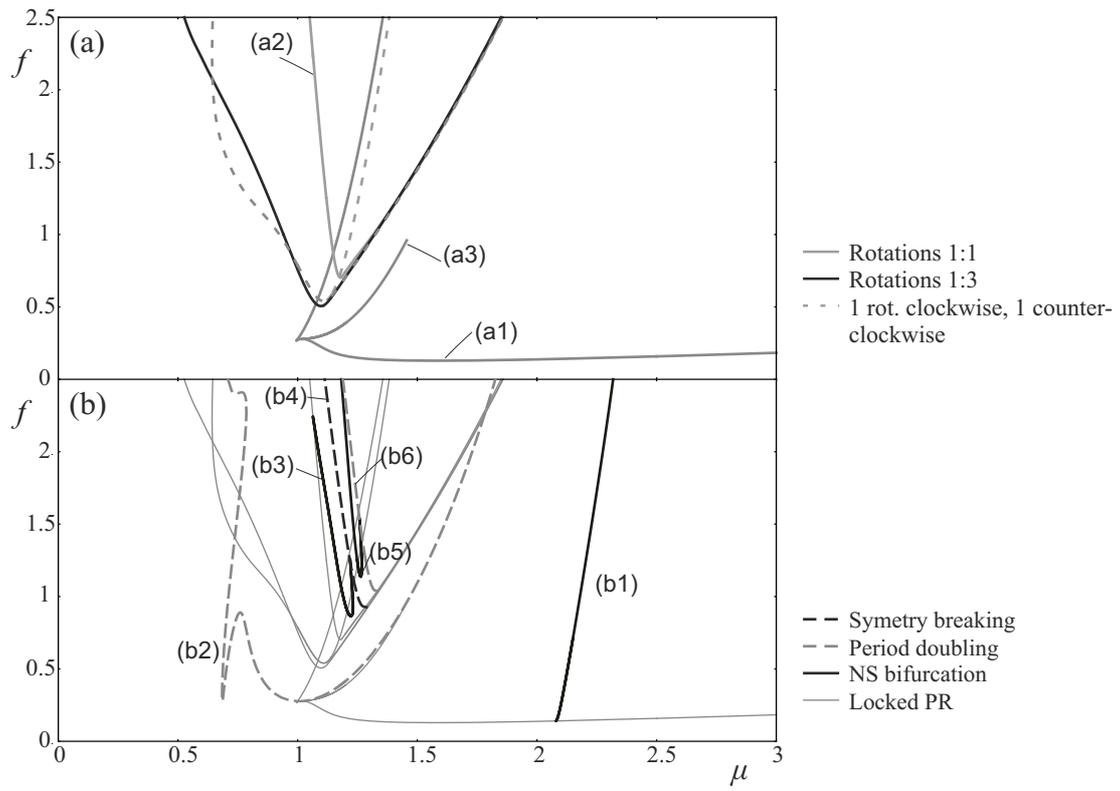} 
\par\end{centering}

\caption{Bifurcation diagram of PR for system (\ref{eq:row bez}) in two parameter
space ($f$,$\mu$). Lines show bifurcations to PR with different
locking ratios (a) and their destruction (b).\label{fig:Bifurcation_oscillating}}
\end{figure}

\subsection{Rotation in plane $(f,\,\mu)$}

In Fig. \ref{fig:Bifurcation_oscillating}(a) we show the bifurcation
diagram of rotations in two-parameter $\left(f,\,\mu\right)$ plane.
To observe a rotational solutions pendulum has to undergo a global
heteroclinic bifurcation. It means that this class of solutions could
not be approached by following a steady state or PO. For continuation,
one has to start from an integrated PR orbit. Two kinds of PR can
occur in a considered system similarly to parametrically forced pendulum.
One with a constant rotational motion in one direction are called
a pure rotations \cite{Clifford1995191} and second one with change
of rotation direction are termed oscillations-rotations \cite{szemplinska}.
All lines presented in Fig. \ref{fig:Bifurcation_oscillating}(a) are saddle-node bifurcations of PR
\cite{Clifford1995191,Xu20051537}. We observe the following resonant
continuous PR: three ranges of the $1:1$ locked PR in clockwise and
counter-clockwise directions (solid grey), the $1:3$ in clockwise
and counter-clockwise directions (solid black line). Bifurcation curves
of both clockwise and counter-clockwise directions overlap due to
the symmetry $\gamma\to-\gamma$. The third state is a repeated sequence
of oscillations-rotations: one rotation in clockwise and one in counter-clockwise
direction ($1L1R$). In Fig. \ref{fig:Bifurcation_oscillating}(b)
we show the bifurcations which destroy PR. The $1:1$ rotational motion
(solid grey line (a1)) is stable in the large range of parameters,
its stability is bounded by the Neimark-Saker bifurcation (see black
line (b1)) in the right part of Fig. \ref{fig:Bifurcation_oscillating}(b)).
The second $1:1$ area is bounded by the solid gray line (a2), with
increasing parameters we observe the Neimark-Saker bifurcation, which
destabilizes this resonance (curve (b5)). The last $1:1$ PR (line
(a3)) is bounded by the period doubling bifurcation curve (curve (b2)).
The $1L1R$ PR goes through the symmetry breaking pitchfork bifurcation
(curve (b4)) and the period doubling bifurcation (curve (b6)) finally
reaches the chaotic attractor in the period doubling cascades (the
detailed description in Fig. \ref{fig:Continuation-in-one}(d)).
The stability of $1:3$ PR is bounded, on right side, by the Neimark-Saker
bifurcation (curve (b3)), where we observe emergence of quasiperiodic
motion, and further transition to chaotic attractor via torus breakdown
with increasing of the amplitude $f$. On the left side this resonance
disappear in period doubling bifurcation (line (b2)).

\subsection{One parameter continuation}

\begin{figure}
\begin{centering}
\includegraphics{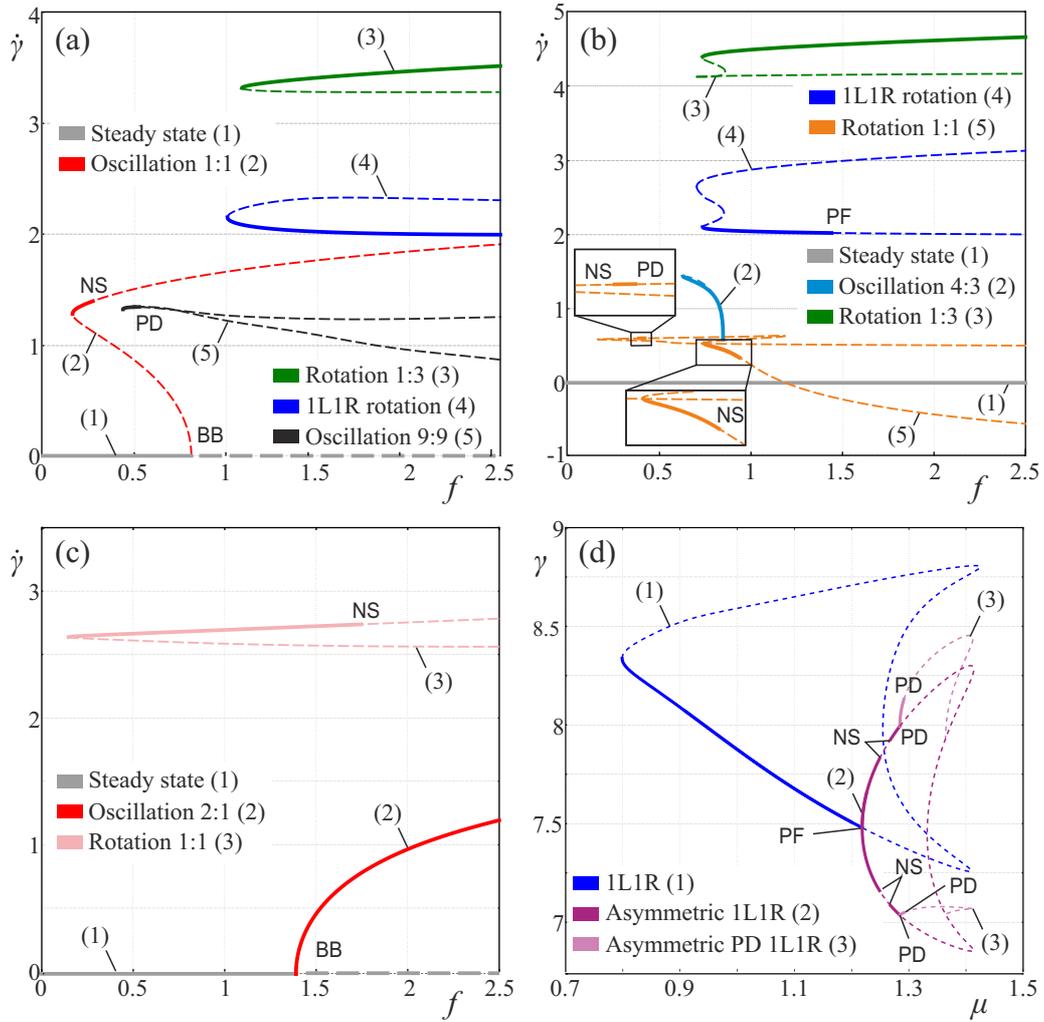} 
\par\end{centering}

\caption{(colour online) Continuations of periodic solutions in one parameter
$f$ for $\mu=0.92$ (a), $\mu=1.2$ (b), and$\mu=2.25$ (c). (d)
shows period doubling route to chaos starting from $1L1R$ rotational
solution for fixed $f=1.25$ and variable $\mu$. Solid and dashed
lines indicate the stable and unstable periodic solution respectively.
Abbreviations: BB - branching bifurcation, PD - period doubling bifurcation,
NS - Neimark-Saker bifurcation, PF - symmetry breaking pitchfork bifurcation.
Other changes of the stability take place through the saddle-node
bifurcations. \label{fig:Continuation-in-one}}
\end{figure}

In this section, we present one-parameter continuations of periodic
solutions (PO as well as PR) versus the amplitude $f$ or frequency
$\mu$ of excitation. We follow periodic solutions emerging in bifurcations
presented in the previous sections. In Fig. \ref{fig:Continuation-in-one}(a-c)
we plot maximum velocity $\dot{\gamma}$ while in Fig. \ref{fig:Continuation-in-one}(d)
maximum angular position $\gamma$. PO and PR presented in Fig.
\ref{fig:Continuation-in-one}(a) are calculated for the frequency
of excitation fixed to $\mu=0.92$ and $f$ is varied. When $f\in(0.0,\,0.812)$
we observe the oscillation of the Duffing oscillator while the pendulum
is in the lower equilibrium position (line (1)). At the end of this
interval this state looses its stability in the subcritical branching
bifurcation and unstable PO is created. For $f=0.162$ the pendulum
oscillations stabilize via saddle-node bifurcation and further destabilize
at $f=0.274$ through the Neimark-Saker bifurcation (curve (2)). For
this value of the frequency $\mu$ there appear rotational solutions.
The line (3) corresponds to rotational resonance $1:3$ and the line
(4) indicates $1L1R$ PR, both solutions appear through the saddle-node
bifurcations for $f=1.11$ and for $f=1.0$ respectively. Line (5)
shows $9:9$ resonance stable in range $f\in\left(0.49,\,0.51\right)$
(this locked solution have small area of existence, and we do not
show it in two-dimensional bifurcations diagram). 

In Fig. \ref{fig:Continuation-in-one}(b) the frequency is fixed
as $\mu=1.2$. We observe a stable steady state of the pendulum in
the whole range of $f$ (line (1)). In this case, the dynamics is
reduced to the motion of the forced Duffing oscillator. We also observe
the $4:3$ locked oscillation for $f\in\left(0.628,\,0.847\right)$
(line (2)), which corresponds to the loop in two dimensional plot
(see right side of $4:3$ resonance in Fig. \ref{fig:Bifurcation_oscillating}(a)).
Stable $1:1$ rotations (line (5)) in both directions appear in the
first range in Neimark-Saker and become unstable in period doubling
bifurcation ( $f\in\left(0.39,\,0.40\right)$). In the second range
$1:1$ PR stabilize by the saddle-node bifurcation and become unstable
through the Neimark-Saker bifurcation for $f\in\left(0.743,\,0.938\right)$.
Last two curves are PR: $1:3$ rotation is stable from $f=0.733$
(line (3)), and $1L1R$ motion is stable in range $f\in\left(0.736,\,1.44\right)$
(line (4)).This last $1L1R$ PR disappears in symmetry breaking pitchfork
bifurcation; the detailed analysis of this PR and its bifurcations
is presented in Fig. \ref{fig:Continuation-in-one}(d). 

In Fig. \ref{fig:Continuation-in-one}(c) we show the periodic solutions
for $\mu=2.25$. One can observe $2:1$ resonant PO (line (2)), which
emerges at the branching bifurcation ($f=1.39$) from stable steady
state of the pendulum (line (1)) and stable $1:1$ rotation (line
(3)) which starts at the saddle-node bifurcation and terminates in
the Neimark-Saker bifurcations at the ends of the interval $f\in\left(0.147,\,1.75\right)$. 

The last plot (Fig. \ref{fig:Continuation-in-one}(d)) shows a route
to chaos starting from $1L1R$ rotations for fixed value $f=1.25$
and variable $\mu$. For frequency $\mu=0.80$, we observe the saddle-node
bifurcation (on line (1)), when $\mu=1.218$ the PR goes through the
symmetry breaking pitchfork bifurcation and two asymmetric $1L1R$
rotation appears (line (2)). Further we observe the range of existence
of stable two dimensional quasiperiodic solution (confirmed by integration),
which starts and ends in the Neimark-Saker ($f=1.249$) and the inverse
Neimark-Saker ($f=1.268$) bifurcations (on line (2)). Finally, we
observe the period doubling route to chaos. We show only the first
solution branch with doubled period (line (3)).

\subsection{Influence of non--linearity of spring}

\begin{figure}
\begin{centering}
\includegraphics{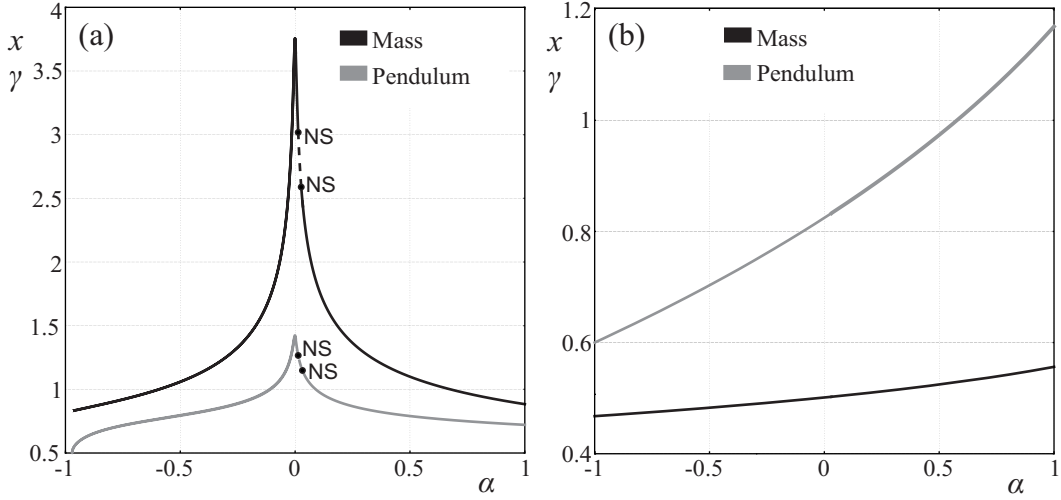} 
\par\end{centering}

\caption{Maximum amplitude of mass (black) and maximum angle of pendulum (gray)
versus non-linearity of spring $\alpha$ in range $\alpha\in\left(-1.0,\,1.0\right)$
for $1:1$ resonance (a) and $2:1$ resonance (b). Characteristic of spring
is changing from ``soft'' to ``hard'' with the increasing of $\alpha$.
Continuous and dashed lines indicate stable and unstable periodic
solutions. Abbreviation NS indicate Neimark-Saker bifurcation. \label{fig:Non_lin_spring}}
\end{figure}

The characteristic of spring have a significant influence on transfer of energy from Duffing oscillator to the pendulum during resonances \cite{kecik2005,Warminski2009612}. In Fig. \ref{fig:Non_lin_spring} we present
these properties for two pairs of parameters: first for $\mu=1.0$
and $f=0.49$ in $1:1$ resonance tongue (a) and second for $\mu=2.0$
and $f=1.02$ in $2:1$ resonance tongue (b). Black and gray colours
indicate respectively the maximum position of mass $M$ -- $x$ and
the maximum angle of the pendulum -- $\gamma$ as functions of the
spring non-linearity $\alpha$. The continuous and dashed lines indicate
respectively stable and unstable PO. It is easy to see that for $1:1$
resonance in the case of $\alpha=0$ linear resonance occurs (Duffing
oscillator is reduced to linear oscillator). For smaller values of
$\alpha$ the amplitude of oscillations is rapidly decreasing for
both pendulum and Duffing oscillator up to $\alpha=-0.97$ where the
motion of the pendulum stops. Similar decreasing of the amplitude
of oscillations appears for positive $\alpha,$but motion terminates
at $\alpha=4.05$. For $\mu=0.05$ we observe the Neimark-Saker bifurcation
followed by the inverse Neimark-Saker bifurcation for $\mu=0.075$.
Between the bifurcation we observe a stable quasiperiodic motion.
In Fig. \ref{fig:Non_lin_spring}(b) one observes a stable $2:1$
PO in the whole range of $\alpha$. When the system is changing from
soft to hard characteristic of spring amplitudes of the Duffing oscillator
and the pendulum are increasing. We do not observe a resonance of
Duffing system. 

\section{Coexistence of solutions }

\begin{figure*}[p]
\begin{centering}
\includegraphics{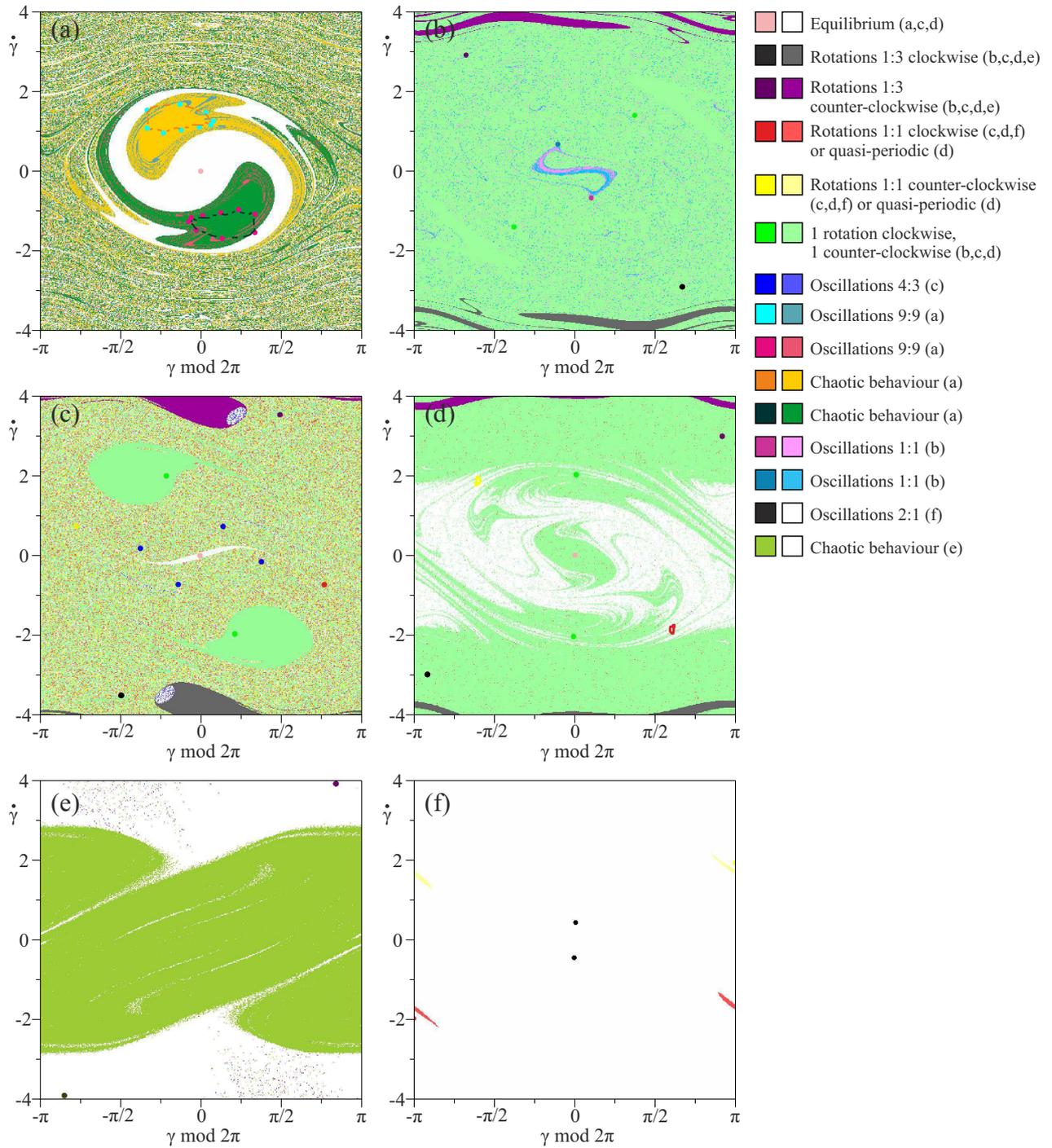} 
\par\end{centering}

\caption{(colour) Basins of attraction in the plane of the pendulum variables
$\left(\gamma,\dot{\gamma}\right)$ for $\mu=0.9$, $f=0.5$ (a);
$\mu=1.03$, $f=1.25$ (b); $\mu=1.2$, $f=0.75$ (c); $\mu=1.229$,
$f=1.0$ (d); $\mu=1.6$, $f=1.95$ (e), and $\mu=2.37$, $f=2.25$
(f). The colours for attractors their basins are shown in legend.\label{fig:Basin-of-attraction}}
\end{figure*}

\begin{figure}
\begin{centering}
\includegraphics{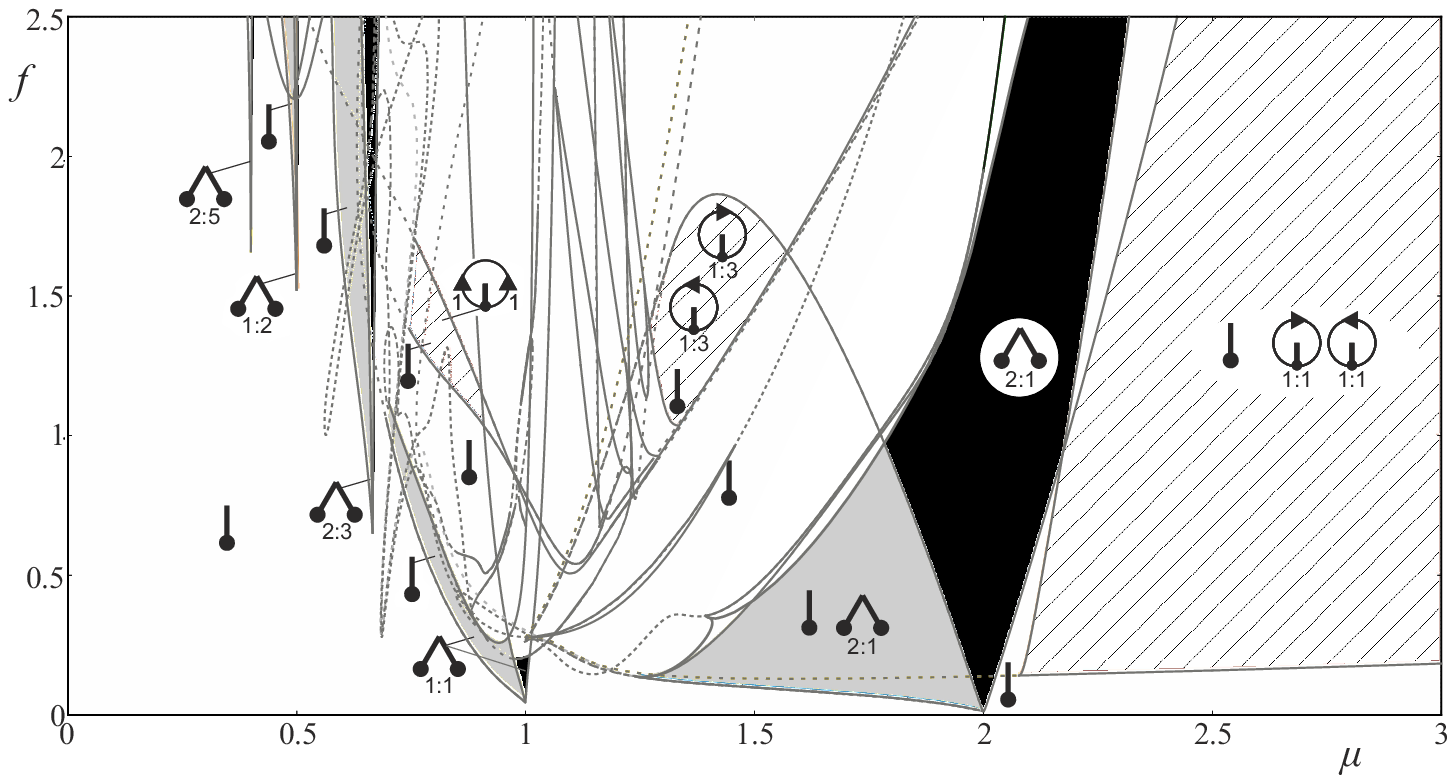} 
\par\end{centering}

\caption{Two-parameter bifurcations diagram in plane $\left(f,\,\mu\right)$
with PO and PR. Black colour indicates one attractor (for locked PO,
excluding $4:3$ resonance, between period doubling bifurcation lines
on the left and right side - edges of the resonance tongues), grey
colour shows two coexisting attractors (the same as for black but with
coexisting stable steady state of the pendulum). In the hatched area
we observe the coexistence of stable rotations (three areas with:
pairs of symmetric $1:1$ or $1:3$ or $1L1R$ PR) and stable steady
state of the pendulum.\label{fig:Two-paramters_colour}}
\end{figure}

As follows from the previous sections, see e.g. Fig. \ref{fig:Continuation-in-one}(a-d),
multiple stable periodic solutions often coexist for the same parameter
values. In this section we illustrate basins of attraction of different
stable periodic solution. Numerical package Dynamics 2 \cite{Nusse_yorke}
is used. We choose six representative sets of parameters $\left(f,\,\mu\right)$
and show the calculated basins in plane $\left(\gamma,\dot{\gamma}\right)$
(angular displacement and velocity of pendulum), where $\gamma$ is
$\textrm{mod}\,2\pi$. Initial conditions of the pendulum are taken
in ranges: $\gamma\in\left(-\pi,\pi\right)$ and $\dot{\gamma}\in\left(4,-4\right)$,
initial conditions of Duffing oscillator are fixed for each plot and
have the following values: $x_{0}=0$, $\dot{x}_{0}=0$ (a,b,d,e),
$x_{0}=2.0$, $\dot{x}_{0}=5.0$ (c, f). Since different initial conditions
of the Duffing oscillator may lead to different attractors in plane
$\left(\gamma,\dot{\gamma}\right)$, the obtained figures show two-dimensional
cross-sections of the four directional phase space (plus the phase
of the perturbation). There no guarantee that for other initial states of Duffing we reach the same set of attractors. As one can see in Fig. \ref{fig:Bifurcation_oscillating}
and Fig. \ref{fig:Two-paramters_colour} most of the periodic solutions
are accumulated in range $\mu\in\left(0.7,\,1.3\right)$, so we calculated
four out of six the basins in this area. In Fig. \ref{fig:Basin-of-attraction}(a)
($\mu=0.9$ and $f=0.5$) we find five attractors: steady state, pair
of period nine motion and chaotic motion which bifurcates from $1:1$
PO. Then in Fig. \ref{fig:Basin-of-attraction}(b) ($\mu=1.03$ and
$f=1.25$) one can observe five attractors. Two of them are symmetric
pairs of rotations ($1:3$ resonance), two corresponds to symmetric
$1:1$ PO and the last one is the $1L1R$ PR. Next plot ($\mu=1.2$
and $f=0.75$) includes seven attractors: two symmetric pairs of PR
($1:1$ and $1:3$), $4:3$ PO, equilibrium of pendulum and $1L1R$
PR. In Fig. \ref{fig:Basin-of-attraction}(d) ($\mu=1.229$ and $f=1.0$)
we detect six solutions, two of them are quasiperiodic bifurcated
from symmetric pair of $1:1$ PR , pair of $1:3$ PR, steady state
of pendulum and $1L1R$ PR. For larger values of excitation we also
find chaotic attractor, e.g. for $\mu=1.6$, $f=1.95$ (see Fig. \ref{fig:Basin-of-attraction}(e)).
Usually the chaotic solution dominates the whole phase space and the
coexisting attractors have small basins of attraction, and such a
situation is also observed in the investigated system (\ref{eq:row bez}).
In the last Fig. \ref{fig:Basin-of-attraction}(f) we show a case
where we do not observe fractal basins of attraction. Most of the
phase space is dominated by period two symmetric PO, only in small
range one can see basins of symmetric pair of $1:1$ PR.

Generally, in area where $1L1R$ PR exists its basin of attraction
dominate in phase space. The $4:3$ resonance could be observed for all
pairs of parameters used in Fig. \ref{fig:Basin-of-attraction}(a-d).
Nevertheless, we see this attractor only in Fig. \ref{fig:Basin-of-attraction}(c),
where different initial conditions of Duffing are used. This is the
evidence that not only sensitivity on initial state is observed for
pendulum but also for Duffing. Varying initial condition of Duffing
moves cross-section of the phase space and changes the set of accessible
attractors. 

Almost all basins of attraction have a fractal structure, so reaching
the given solution is strongly dependent on initial conditions. From
the practical point of view it is important to know the area with
a small number or even one solution \cite{ChudzikPSK11}. In such
ranges one can be sure that the system approaches the expected solution.
In Fig. \ref{fig:Two-paramters_colour} we marked the areas with different
types of attractors: black colour indicates one attractor (four locked
PO, excluding $4:3$ resonance, between two branching bifurcation
lines on the left and right side - the edges of the resonance tongues),
grey colour refers to two coexisting solutions (the same as for black
but with coexisting steady state of the pendulum). In the hatched
area we observe the coexistence of PR (two symmetric pairs of $1:1$
or $1:3$ or $1L1R$ PR) and the steady state of the pendulum. The
largest area with one attractor is a tongue of $2:1$ resonance. For
other resonances ($1:1$, $2:3$, $2:5$, $1:2$) these areas are
small. Especially, it is surprising for $1:1$ where the resonance
tongue is large in the parameter space but only near its edge we do
not find the second attractor. We do not mark the areas where only
Duffing system is oscillating and pendulum is stable equilibrium position.
In this case the dynamics of the system is reduced to the oscillations
of mass $\left(M+m\right)$.

\section{Conclusions}

We present a comprehensive numerical analysis of the forced Duffing
oscillator with the suspended pendulum. We show two dimensional bifurcation
diagrams with the most representative periodic solutions and demonstrate
the bifurcation route to the locked resonances. The linear resonance
of both subsystems is observed for $\mu=1.0$ and around this value
we find complex dynamics with many coexisting attractors, not only
periodic but also quasiperiodic and chaotic ones. In the principal
resonance zone the pendulum oscillations decrease the oscillation
amplitude of the Duffing oscillator so we observe an energy transfer
from the Duffing oscillator to the pendulum. These properties
are observed only for $1:1$ oscillatory resonance, for other locked
solutions (oscillations or rotations) the pendulum oscillations increase
the oscillations amplitude of the Duffing oscillator (the control
and the possible decrease of the amplitude for other resonances will
be considered in future work). Contrary to complex dynamics around
$1:1$ principal resonance in the neighbourhood of $2:1$ parametric
resonance, we find two large ranges in parameters space with only
one attractor ($2:1$ locked oscillations) and symmetric pair of $1:1$
rotations respectively. From the practical point of view (certainty
of reaching the desired attractor) such a situation is very useful
and rare in non--linear system. We also compare the influence of non-linearity
of spring on the amplitude of oscillations For principal $1:1$ resonance,
strong non-linearity (hardening or softening) causes lower amplitude
of oscillations. For $2:1$ parametric resonance only the decrease
of $\alpha$ into the softening direction causes the decrease of the
oscillations amplitude. The existence of fractal basins of attraction
is not surprising for non--linear systems with the attached pendulum.
Hence, we can consider only the probability of reaching the chosen
attractor and never have a certainty where the dynamics of systems
evolves. This is crucial when the pendulum is working as a tune mass
absorber so we show the areas with a low number of coexisting solutions
in the parameters space.

\section*{Acknowledgement}

This work has been supported by the Foundation for Polish Science,
Team Programme under project TEAM/2010/5/5(P.B., P.P. and T.K.), Foundation
for Polish Science, the START fellowship (P.P). and DFG Research Center
MATHEON \textquotedblright{}Mathematics for key technologies\textquotedblright{}
under the project D21 (S.Y.).

\section*{References}

\bibliographystyle{elsarticle-num}

\end{document}